\documentclass[onecollarge]{svjour2}

\smartqed

\usepackage{mathptmx}
\usepackage{graphicx}
\usepackage{amssymb}
\usepackage{amsmath}
\usepackage[latin1]{inputenc}
\usepackage{float}

\journalname{Foundations of Physics}

\begin{document}

\title{On the interpretative essence of the term ``interaction-free measurement'': The role of entanglement}
\author{Renato M. Angelo}

\institute{Renato M. Angelo \at 
  Department of Physics, Federal University of Paraná, P.O. Box 19044, 81531-990, Curitiba, PR, Brazil \\
  Tel.: +55 41 3361-3275 \\
  Fax: +55 41 3361-3418 \\
  \email: renato@fisica.ufpr.br
}

\date{Received: date / Accepted: date}

\maketitle

\begin{abstract}
The polemical term ``interaction-free measurement'' (IFM) is analyzed in its interpretative nature. Two seminal works proposing the term are revisited and their underlying interpretations are assessed. The role played by nonlocal quantum correlations (entanglement) is formally discussed and some controversial conceptions in the original treatments are identified. As a result the term IFM is shown to be consistent neither with the standard interpretation of quantum mechanics nor with the lessons provided by the EPR debate.  
\end{abstract}

\section{Introduction}
\label{intro}

The term {\em interaction-free measurement} (IFM) appeared for the first time in 1981 in a paper by Dicke~\cite{Dicke-81}. Inspired by Renninger's ideas of ``negative result experiments''~\cite{Renninger-60} Dicke concludes that the wave function of a particle may change even in a situation in which a light beam is not scattered when passing through the region where the particle is likely to be found. Since some information bas been obtained in the process an IFM is claimed to occur.

In 1993, Elitzur and Vaidman (EV) invoked the term IFM to nominate a peculiar technique that allows for detecting infinitely fragile objects without destroying them ~\cite{Elitzur-93}. Along the subsequent decade, the term IFM received several criticisms, which make Vaidman return to the subject in an effort to dissipate the confusion generated around the term (see a more recent work by Vaidman~\cite{Vaidman-03} and references therein to track the whole discussion). Despite its unusual features, the technique has received experimental support~\cite{Kwiat-95,Kwiat-99} and a fair bit of discussion~\cite{Ghirardi-05,Sant'Anna-06}.

After this long period it seems, however, not to exist a consensual answer for the central question: {\em Is IFM really interaction-free?} In evident conflict, some authors state: ``Of course IFMs are not really interaction free''~\cite{Potting-00}; ``It is a triviality that in all cases some kind of interaction, describable by a Hamiltonian, plays an important role in the unitary stage of the processes''~\cite{Tamas-98}; and then ``We stress that from the viewpoint of a single event, a successful measurement is {\em completely} interaction-free''~\cite{Kwiat-95}. 

Within this context of conflicting views, it seems natural to suspect that the term IFM may be associated with some specific interpretation of quantum mechanics, thus being strongly dependent on it. In fact, since different interpretations can be adopted for the quantum formalism -- this one not being capable of deciding which is ``the correct'' -- this would explain why a measurement can be understood as {\em being} or {\em not being} interaction-free. In this paper, two original works proposing the term IFM are revisited in the light of the current knowledge about quantum correlations, which can be rigorously calculated from the quantum formalism. As a result, it is shown that the term IFM is in conflict with both the knowledge produced by the EPR debate~\cite{EPR,Bell,Aspect-82} and the standard interpretation of quantum mechanics. Our analysis does not prohibits the use of the term IFM, but points out the need for attaching the adopted interpretation in order to prevent the propagation of conceptual misunderstandings.

\section{Revisiting Dicke's null experiment}
\label{dicke}

\begin{figure}[ht]
\begin{center}
\includegraphics[scale=0.4,angle=0]{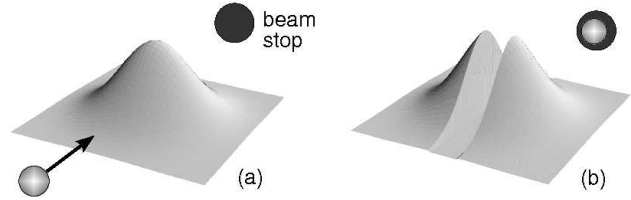}
\end{center}
\caption{Simplified version of the thought experiment proposed by Dicke. (a) A pulse of radiation (the sphere) is focused on a region where an ion is initially prepared in a Gaussian state. (b) If not scattered by the trapped particle, the pulse is collected in a beam stop, none of the detectors clicks (a null result), and, according to Dicke's analysis, the particle wave function suffers a kind of erasing even though no apparent interaction has taken place.} \label{fig1}
\end{figure}

In his paper of 1981~\cite{Dicke-81}, Dicke considers a trapped particle (an ion) initially prepared in the ground state of the trapping potential (see the Gaussian-like distribution in Fig.~\ref{fig1}(a)).  An intense pulse of radiation, represented by the sphere in Fig.~\ref{fig1}, moves towards the region where the particle is trapped. Detectors are placed all around the trap (ideally a $4\pi$ solid angle counter) in order to collect any scattered photon, and a beam stop, placed behind the trap, absorbs all undeviated light. Only two outputs are expected in this thought experiment: (i) photons are counted in the detectors, implying that the particle was in the path of the beam, or (ii) no photon is counted (a null result), meaning that the particle was out of the beam path.

According to Dicke, the paradoxical situation occurs in the second case: When no photon is scattered, the particle is certainly out of the region illuminated by the beam, so that $\psi=0$ in this region (see Fig.~\ref{fig1}(b)). As a consequence, the particle wave function changes even without any apparent interaction between the photon and the particle.

Although the argument seems to be somewhat compelling it actually is controversial, or in the best case, deeply dependent on the interpretation. Let me give some arguments supporting this statement. Firstly, notice that the claimed ``wave function erasing'' would produce discontinuities in the particle wave function and thus the momentum expectation value, $\langle \mathbf{p}\rangle=-i\hbar\int d³\mathbf{r}\psi^*(\mathbf{r},t)\nabla\psi(\mathbf{r},t)$, would not be well behaved. Secondly, Dicke analyzed a collision between {\em physical corpuscles} (the photons in the beam) with the square modulus of the {\em wave function} of the trapped particle, as is depicted by Fig.~\ref{fig1}. This is a rather unusual description which mixes physically different entities. It would be much more reasonable, and more consistent with quantum formalism, to investigate the collision among wave entities: A particle wave function and the wave functions of each photon of the beam (or a unique wave function describing the global set of correlated photons in the beam). In this fully quantum-mechanical treatment, things would be much more difficult to describe heuristically. Before the light pulse enters the region where the particle is likely to be found, the global vector state of the system might be written as a tensorial product, $|\Psi_0\rangle=|\psi_{\textrm{particle}}\rangle\otimes|\psi_{\textrm{beam}}\rangle$. After passing through the interaction region, the system would become entangled and the quantum state would assume a general form such as $|\Psi_t\rangle=|\psi_{\textrm{particle $+$ beam}}\rangle$ which cannot be factorized in a tensorial product of individual vector states corresponding to each corpuscle of the system. Then, in order to graphically represent the wave collisions we could calculate, in the best case, density matrices for each individual corpuscle by tracing over the undesirable degrees of freedom. Certainly, in this case, Dicke could not predict that kind of erasing in the particle wave function as represented in Fig.~\ref{fig1}(b).

Dicke's interpretation seems to derive from a conception which tries to associate the wave function with an individual realization of the experiment. In fact, this idea already appears in the motivating question proposed in his paper~\cite{Dicke-81}: ``Is the particle located in the light beam or not?'' In general, this question cannot be answered by the quantum theory, which is able to predict only the {\em probability} of finding the particle in a run of several similarly prepared experiments ({\em ensemble}). When a light pulse is sent towards the trap we have only a single realization of the experiment. If in a realization no photon is scattered, in another, one may be. Quantum results cannot be related with the first or the second, but with all realizations composing the statistics of the experiment. The connection between quantum predictions with the experimental realization of statistical ensembles consists a crux of both the standard and the statistical interpretation~\cite{Ballentine} of quantum mechanics. (In agreement with such ideas, recall that it is not possible at all to identify a diffraction pattern on a screen when only one electron is sent through the single slit!) Therefore, we may conclude that the erasing in the wave function as claimed by Dicke is misleading.

Further, in trying to corroborate his own claims Dicke applies formal measurement theory to analyze the null experiment. Using an initial state that mixes spatial functions with vector states in an unclear way and a non-normalized projection operator (in conflict with the measurement postulate of the quantum mechanics), the author concludes that the center-of-mass wave function of the target particle changes even when no scattering occurs. According to the quantum formalism this is not expected at all since no coupling between the center of mass and the others degrees of freedom has been taken into account in the Hamiltonian of the system. This point can be assessed by means of an analysis carried out on a simplified version of Diche's proposal.  Let us replace the light pulse with a large {\em probe} particle (``particle $P$'') and the trapped ion with a free {\em target} particle (``particle $T$''). The problem is now reduced to a local collision between two distinguishable quantum particles. Initially, the particles are far apart and the joint quantum state describing the system can be written as $|\psi_{\textrm{before}}\rangle=|\textrm{free}\rangle_P\otimes|\textrm{free}\rangle_T$, where $|\textrm{free}\rangle_{P(T)}$ denotes the vector state of the particle $P\,(T)$ before entering the interaction region. These states may be thought of as increasing variances distributions~\cite{Cohen} in the configuration space. According to the quantum formalism, the time-evolved vector state would be written as
\begin{eqnarray}\label{psi_after}
|\psi_{\textrm{after}}\rangle=\alpha
|\textrm{free}\rangle_P\otimes|\textrm{free}\rangle_T+\beta|\textrm{scatt}\rangle_P\otimes|\textrm{scatt}\rangle_T,
\end{eqnarray}
where $|\alpha|^2+|\beta|^2=1$. Here $|\textrm{scatt}\rangle_{P\,(T)}$ denotes the scattering of the particle $P\,(T)$.

Notice by \eqref{psi_after} that when no scattering is detected (a null result) both particles must remain in their original free-particle quantum states, with no alteration in their respective mean momenta. Since this simple model retains the essence of the problem investigated by Dicke, one may conclude that no ``erasing'' in the wave function of the particle $T$ is expected. In such a case, the target wave function must remain unchanged. Thus, we have arrived at an important point: The nonexistence of the wave function erasing is formally predicted by the quantum formalism without any appeal for particular interpretations. This reveals a weakness in Dicke's description of the experiment. In principle, one might object that the simple model used does not really fit Dicke's proposal, but the application of Dicke's reasoning would erroneously predicts that kind of erasing in this simple model as well. 

Lastly, the initial question may be reformulated as follows: Does some interaction really take place when a null result is obtained in a {\em single} realization of the experiment? The state given by \eqref{psi_after} induces us to interpret that a null result implies absence of any interaction. In fact, in this case the particles keep moving as free particles. This question will be discussed in details in the next section, once this is a central point also in the EV experiment.

\section{Revisiting the EV proposal}
\label{EV}

\begin{figure}[ht]
\begin{center}
\includegraphics[scale=0.35,angle=0]{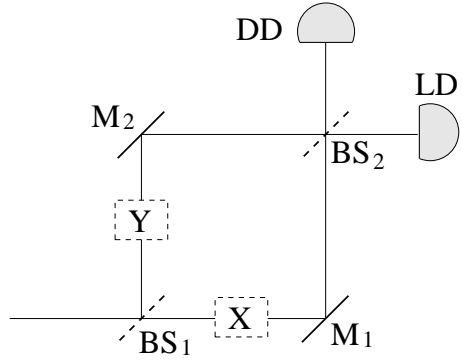}
\end{center}
\caption{Generalization of the EV experiment. An object is initially prepared in region $X$, $Y$, or in a quantum superposition of $X$ and $Y$.}
\label{fig2}
\end{figure}

The EV scheme is based on the Mach-Zehnder interferometer (see Fig.~\ref{fig2}). Individual photons are sent horizontally towards the first beam splitter (BS$_1$) with a transmission coefficient 1/2. With no obstacles in the arms the transmitted and reflected parts of the photon wave are reflected, respectively, by mirrors M$_1$ and M$_2$, being afterwards reunited at the beam splitter BS$_2$, whose transmission coefficient is 1/2 too. Two photon detectors, LD (``light detector'') and DD (``dark detector''), are positioned according to Fig.~\ref{fig2}. The setup geometry is such that no photon arrives at DD (destructive interference) when the arms are empty. On the other hand, when an object is at some arm, say in the region $X$, the condition of destructive interference is no longer satisfied and DD collects the photon with probability 1/4. Then, when in some realization of the experiment a single photon is detected at DD, we can know for sure that there is something in one of the arms of the interferometer. Since in this case the photon has not been absorbed by the object, thus apparently not interacting with it, this is called an interaction-free measurement.

Note at first that knowing that an object is inside the interferometer does not provide accurate information about the vector state of the object. In fact, any state $|\Psi\rangle$ yielding a wave function $\Psi(\mathbf{r})=\langle\mathbf{r}|\Psi\rangle$ that is finite within the interferometer arms may lead to a click in DD in some realization of the experiment. However, this information alone does not allow for determining the state $|\Psi\rangle$ precisely. In this sense one may object whether the term {\em measurement} is indeed appropriate. 

The question whether IFM is really interaction-free is, of course, the most subtle. In order to investigate this point we review the EV calculations with some convenient generalizations. Consider that before the photon enters the interferometer the quantum state of the system ``photon $+$ object'' is given by
\begin{eqnarray}\label{psi0}
|\psi_0\rangle=|1_x0_y\rangle\otimes\Big(\alpha|GX\rangle+\beta|GY\rangle
\Big),
\end{eqnarray}
with $|\alpha|^2+|\beta|^2=1$. The state $|1_x0_y\rangle$ indicates the situation in which there is one photon in the $x$ direction (horizontal paths in Fig.~\ref{fig2}) and none in the $y$ direction (vertical paths). The object is assumed to have been initially prepared in its lowest level of energy $(G)$ and in a quantum superposition of regions $X$ and $Y$.

After the photon crosses the beam-splitter BS$_1$ the quantum state reads
\begin{eqnarray}\label{psi1}
|\psi_1\rangle=\left(\frac{|1_x0_y\rangle+i|0_x1_y\rangle}{\sqrt{2}}\right)\otimes\Big(\alpha|GX\rangle+\beta|GY\rangle \Big).
\end{eqnarray}
The phase $i$ must be added after each reflection by a beam-splitter. Next, the photon visits the region where the object is likely to be in. Following the EV approach we obtain
\begin{eqnarray}\label{psi2}
|\psi_2\rangle=|\psi_{\textrm{coll}}\rangle+\frac{\beta}{\sqrt{2}}|1_x0_y\rangle\otimes|GY\rangle +\frac{i\,\alpha}{\sqrt{2}}|0_x1_y\rangle\otimes|GX\rangle,
\end{eqnarray}
where
\begin{eqnarray}\label{psi_coll}
|\psi_{\textrm{coll}}\rangle\equiv|0_x0_y\rangle\otimes\left(\frac{\alpha|EX\rangle+i\,\beta|EY\rangle}{\sqrt{2}}\right).
\end{eqnarray}
The state $|\psi_{\textrm{coll}}\rangle$ describes the portion of the Hilbert space accounting for the collision between the corpuscles: The photon is absorbed and the object reaches its excited state $|E\rangle$. Note that the collision is supposed to occur only when the photon and the object are at the same arm, this being an indicative of the assumption of a {\em local interaction}. 

The state \eqref{psi2} presents a remarkable difference from the one deduced originally by EV~\cite{Elitzur-93}, namely, it displays nonlocal correlations (entanglement). It is worth emphasizing that this difference does not derive from the generalization that has been proposed here. In fact, entanglement is also there if we set $\alpha=0$ and $\beta=1$ (EV case), as we shall see next. The difference comes from the fact that EV originally described the experiment in terms of a single Hilbert space, the one associated with the photon.

After the reflection of the photon by the mirrors, the global quantum state reads
\begin{eqnarray}\label{psi3}
|\psi_3\rangle=|\psi_{\textrm{coll}}\rangle
-\frac{\beta}{\sqrt{2}}|0_x1_y\rangle\otimes|GY\rangle-\frac{i\,\alpha}{\sqrt{2}}|1_x0_y\rangle\otimes|GX\rangle.
\end{eqnarray}
Finally, after the second beam-splitter we have
\begin{eqnarray}\label{psifinal}
|\psi_{\textrm{final}}\rangle=|\psi_{\textrm{coll}}\rangle-
i\frac{|1_x0_y\rangle}{2}\otimes\Big(\alpha|GX\rangle+\beta|GY\rangle\Big)+\frac{|0_x1_y\rangle}{2}\otimes\Big(
\alpha|GX\rangle-\beta|GY\rangle\Big).
\end{eqnarray}
As mentioned above, EV results are reproduced by setting $\alpha=0$ and $\beta=1$:
\begin{eqnarray}\label{psiEV}
|\psi_{\textrm{EV}}\rangle=|\psi_{\textrm{coll}}\rangle-
\frac{i}{2}|1_x0_y\rangle\otimes|GY\rangle -
\frac{1}{2}|0_x1_y\rangle\otimes|GY\rangle.
\end{eqnarray}

Note by \eqref{psiEV} that when DD clicks the state of the object collapses to $|GY\rangle$ -- the initial state of the object. In this case, the term ``interaction-free'' seems to be appropriate, since neither the photon has been absorbed by the object nor the object state has changed. However, in \eqref{psifinal} we identify a more interesting situation which will help us to understand why this ``adequacy'' is only apparent. If DD clicks, the object state collapses to $|\textrm{obj}_{\textrm{final}}\rangle=\alpha|GX\rangle-\beta|GY\rangle$, which is rather different from the initial state $|\textrm{obj}_{\textrm{initial}}\rangle=\alpha|GX\rangle+\beta|GY\rangle$. In fact, the correlation between these two states, which is given by
\begin{eqnarray}\label{C}
\mathbb{C}\equiv  \left|\langle \textrm{obj}_{\textrm{final}}|\textrm{obj}_{\textrm{initial}}\rangle\right|^2=\left(|\alpha|^2-|\beta|^2\right)^2,
\end{eqnarray}
shows that while there is no change in the state of the object in the EV experiment ($\mathbb{C}=1$) those states turn out to be orthogonal $(\mathbb{C}=0)$ when $|\alpha|=|\beta|=1/\sqrt{2}$. This refers to a surprising situation in which the state of the object changes even when no photon is absorbed by the object. Is it possible to maintain the term ``interaction-free'' in referring to this situation? As far as I can see, \eqref{C} obligates us to give a negative answer to this question, since the object possesses no kind of self-dynamics. That is, the state of the object can change only by means of an external influence, which in this case is provided by {\em some} interaction with the photon. The coupling operator $\hat{H}_{\textrm{int}}$, which has to be present in the global Hamiltonian in order to produce the inseparability of the state \eqref{psifinal}, refers to a local potential, as mentioned above. Then, the puzzling question is to identify (if possible) such a ``nonlocal interaction'' which is capable of changing the state of the object, making $\mathbb{C}<1$, even when there is no local collision between the object and the photon.

Our answer for the problem is given in terms of nonlocal quantum resources. The main ingredient in the EV problem is the association of a local interaction $\hat{H}_{\textrm{int}}$, describing an energy exchange, with nonlocal quantum states, which are introduced in the setup by the first beam-splitter. The combination of these elements produces the nonlocal correlations that play central role in several controversial issues underlying the quantum phenomena, as for instance in the EPR debate~\cite{EPR,Bell,Aspect-82}. In fact, as we shall see bellow, entanglement is the angular stone of the IFM.

\begin{figure}[ht]
\vspace{0.5cm}
\begin{center}
\includegraphics[scale=0.35,angle=0]{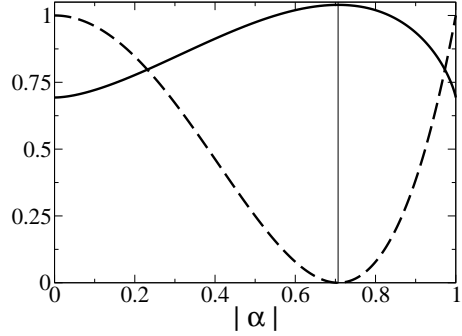}
\end{center}
\caption{Degree of entanglement $\mathcal{E}(|\psi_{\textrm{final}}\rangle)$
(solid line) and correlation $\mathbb{C}$ (dashed line) as a
function of $|\alpha|$. The maximum (minimum) value for the entanglement
(correlation) occurs for $|\alpha|=|\beta|=1/\sqrt{2}$ (vertical
line).} \label{fig3}
\end{figure}

In globally pure bipartite systems, the degree of entanglement is quantified by the von Neumman entropy, which is defined as $\mathcal{E}(|\psi\rangle)=-\textrm{Tr}_{1} (\hat{\rho}_{1}\ln \hat{\rho}_{1})$, with $\hat{\rho}_{1}=\textrm{Tr}_{2}(|\psi\rangle\langle\psi|)$. The subindices 1 and 2 denote either of the subsystems ``photon'' and ``object.'' From \eqref{psifinal} we can obtain and diagonalize $\hat{\rho}_{1}$ and then calculate the entanglement via $\mathcal{E}(|\psi\rangle)=-\sum_{n} \lambda_{n}\ln\lambda_{n}$, where $\lambda_{n}$ is the $n$-th eigenvalue of $\hat{\rho}_{1}$. The result reads
\begin{eqnarray}\label{E}
\mathcal{E}(|\psi_{\textrm{final}}\rangle)=\ln 2-|\alpha|^2\ln |\alpha|-\left(1-|\alpha|^2\right)\ln\sqrt{1-|\alpha|^2}.
\end{eqnarray}
The relation $|\alpha|^2+|\beta|^2=1$ has been used to eliminate $|\beta|$. In Fig.~\ref{fig3} the degree of entanglement \eqref{E} and the correlation function \eqref{C} are plotted in terms of the modulus of the coefficient $\alpha$. 

A first important point in the analysis concerns the conception, grounded on the standard interpretation and generally accepted by researchers in the field of quantum information, that it is impossible to entangle two systems without making them interact, at least indirectly~\cite{Pan-98}. In the formalism, the inseparability of the global state is dynamically induced by the operator $\hat{H}_{\textrm{int}}$. According to this interpretation, since entanglement is present even in the EV case ($|\alpha|=0$), we have to accept that an interaction takes place during the photon dynamics. Also remarkable is the fact shown by Fig.~\ref{fig3} that the lowest correlation, and therefore the most flagrant fail in the ``interaction-free'' interpretation, occurs precisely when the entanglement reaches a maximum. In this case, DD clicks (the photon is not absorbed by the object) but the state of the object changes dramatically ($\mathbb{C}=0$). This result emphasizes the role played by the nonlocal correlations in effectively inducing that kind of ``nonlocal interaction.''

These ideas can be corroborated by a hidden-variables theory, namely, the Bohmian theory~\cite{Bohm}. Consider a EV-like experiment in which the photon is replaced with a probe particle of mass $m$ (see Marton's experimental proposal~\cite{Marton-52} for a Mach-Zehnder interferometer with particles). According to the Bohmian theory, even when the object is not on the path of the probe particle -- this corresponding to a situation of null Newtonian potential -- there is a quantum potential associated with the system wave function, namely, $U(\mathbf{r})\equiv-(\hbar^2/2m)(\nabla^2|\psi|)/|\psi|$. The nonlocality introduced by the beam-splitter on the wave function attributes an underlying nonlocal character to the Bohmian potential. As a consequence, a ``nonlocal potential'' will always be present influencing the Bohmian trajectory regardless the arm the probe particle ``chooses'' to cross the interferometer. Consequently, the term ``interaction-free'' is not admissible by this interpretation as well.

Now, let us focus strictly on the EV case, where the situation is such that $\mathbb{C}=1$ and $\mathcal{E}(|\psi_{EV}\rangle)=\ln 2$. Firstly, it is important to realize that while the correlation $\mathbb{C}$ is an information connecting the initial state of the object (``the before'') with its final state collapsed by the detection (``the after''), the degree of entanglement gives us information about the state of the global system when the photon is inside the interferometer (``the during''). This is a major subtlety in the EV experiment: Even though after the detection the final state of the object displays no change, during the passage of the photon by the interferometer we cannot state, according to the standard interpretation, that no interaction has taken place. In fact, according to the EPR lessons, it is not possible to interpret the state \eqref{psiEV} as a {\em predefined} combination of two local realities such as ``photon absorbed $+$ object excited'' and ``photon in the $X$ ($Y$) path $+$ object in the ground state.'' The same remarks apply to the state \eqref{psi_after} of Dicke's experiment. In these cases a more subtle interpretation is required which conceives these pure states as nonlocal correlated superpositions of those orthogonal realities. The EPR debate~\cite{EPR,Bell,Aspect-82} teaches us that we cannot state that the branch ``free'' (without interaction) or ``scattered'' (with interaction) has been chosen by the particles {\em a priori}, at the beginning of the experiment. Rather, both branches, ``with'' and ``without'' interaction, coexist in a flagrant nonclassical physical state. In Schrödinger's words, this superposition of nonlocal correlations is  ``the characteristic trait of quantum mechanics, the one that enforces its entire departure from classical lines of thought''~\cite{Schrodinger-35}. 

\begin{figure}[ht]
\vspace{0.5cm}
\begin{center}
\includegraphics[scale=0.35,angle=0]{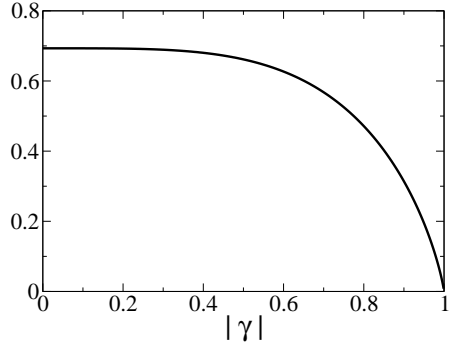}
\end{center}
\caption{Degree of entanglement $\mathcal{E}$ as a function of $|\gamma|$ for the state $|\psi_{\textrm{final}}\rangle$ given by \eqref{psi_finalR2}.}
\label{fig4}
\end{figure}

In fact, entanglement is in the roots of the IFM experiment and hence interpretations which assumes that only one branch occurred are in conflict with the lessons provided by the EPR debate. In order to prove the vital importance of entanglement for the IFM we slightly change the EV model. Inspired by \eqref{psi_after} of Dicke's experiment we consider the following ``interaction'' between the photon and the object at the step 2:
\begin{eqnarray}\label{R2}
|0_x1_y\rangle\otimes|GY\rangle \quad \to \quad \gamma |0_x1_y\rangle\otimes|GY\rangle+\delta|0_x0_y\rangle\otimes|EY\rangle,
\end{eqnarray}
where $|\gamma|^2+|\delta|^2=1$. For $\gamma=0$ and $\delta=1$ we obtain the EV approach. Prescription \eqref{R2} tells us that an interaction operator $\hat{H}_{\textrm{int}}$ has to be present in the Hamiltonian of the system. Of course, depending on the interpretation, for $|\delta|\neq 0$ it is still possible to defend that there is a probability $|\gamma|^2$ of occurring no interaction in a single realization of the experiment. Nevertheless, this interpretation will conflicts the standard interpretation (supported by the EPR debate) since entanglement is always present for $|\delta|\neq 0$. Using the prescription \eqref{R2} and the EV initial state $|\psi_0\rangle=|1_x0_y\rangle\otimes|GY\rangle$ we may reproduce straightforwardly all the calculations to show that
\begin{eqnarray}\label{psi_finalR2}
|\psi_{\textrm{final}}\rangle=-\frac{(1-\gamma)}{2}|0_x1_y\rangle\otimes|GY\rangle-\frac{i(1+\gamma)}{2}|1_x0_y\rangle\otimes|GY\rangle+\frac{i\delta}{\sqrt{2}}|0_x0_y\rangle\otimes|EY\rangle.
\end{eqnarray}
Notice that for $\gamma=1$ the condition of calibration of the EV experiment is recovered. The entanglement degree of the state \eqref{psi_finalR2} is given by
\begin{eqnarray}\label{EfinalR2}
\mathcal{E}(|\psi_{\textrm{final}}\rangle)=\ln 2-\left(1+|\gamma|^2 \right)\ln\sqrt{1+|\gamma|^2}-\left(1-|\gamma|^2\right)\ln\sqrt{1-|\gamma|^2}.
\end{eqnarray}
This is a monotonically decreasing function of $|\gamma|$, as can be seen from Fig. \ref{fig4}.
Equations \eqref{psi_finalR2}, \eqref{EfinalR2}, and the Fig. \ref{fig4} demonstrate our main point. Since the case $\gamma=e^{i\theta}$ (which also produces $\mathcal{E}=0$) has to be disregarded for not yielding the correct calibration of the EV experiment we then conclude that entanglement vanishes only for $\gamma=1$. However, this is precisely the value which prevents detector DD to click (see \eqref{psi_finalR2}). For any other $\gamma$ satisfying $|\gamma|^2<1$, detector DD is allowed to click but entanglement will be present. Therefore, the EV IFM cannot exist without entanglement. Ironically enough, however, the very nonlocal correlations that guarantee the success of the EV experiment prohibit us to interpret the experiment as ``interaction-free.''

Therefore, it seems correct to state that the applicability of the term ``interaction-free'' is strongly dependent on the interpretation adopted for the quantum mechanics and hence, as a matter of consistence, the interpretation should be attached to the term. As far I can see, the EV interpretation is corroborated by the result shown for the correlation $\mathbb{C}$, which allows for the assumption that only one branch exists in each realization of the experiment. As pointed out above, however, there exist alternative interpretations -- supported by nonlocal quantum resources -- which require nonclassical conceptions. For these ones, both branches coexist and the term IFM is no longer admissible. The quantum formalism, which democratically furnishes both $\mathbb{C}$ and $\mathcal{E}$, cannot decide which is the best view. As a consequence, it becomes clear that the polemic around the IFM is supported by the degree of freedom provided by the interpretative essence of the term.

\section{Conclusions}

In this contribution the term ``interaction-free measurement'' is critically revisited in the light of formal results provided by the quantum formalism for the entanglement, which appears to be inevitably present in IFMs. Firstly, Dicke's paper~\cite{Dicke-81} is assessed and some controversial conceptions adopted by the author in interpreting a null result experiment are pointed out. Then, the EV work~\cite{Elitzur-93} is investigated and alternative interpretations for the IFM puzzle in terms of nonlocal quantum resources are proposed.

Both works do not consider the central role played by the entanglement in the dynamics of their thought experiments. According to the standard interpretation, entangled quantum states such as \eqref{psi_after}, \eqref{psifinal}, \eqref{psiEV}, and \eqref{psi_finalR2} cannot be interpreted as classical combinations of two distinguishable realities such as ``with interaction'' {\em or} ``without interaction.'' Rather, as has been shown by the EPR debate~\cite{EPR,Bell,Aspect-82}, a more intricate interpretation is required which conceives such states as two distinguishable realities coexisting simultaneously in a flagrant nonclassical way. In this case, the correct conception is expected to be ``with interaction'' {\em and} ``without interaction,'' thus not being possible to assert that the system occupies only one of the two branches. \footnote{Note that the same question underlies the Schrödinger cat paradox, which is expressed in terms of a superposition comprising ``nucleus in its ground state; cat dead'' {\em and} ``nucleus excited; cat alive.'' Recall that the classical interpretation ``nucleus in its ground state; cat dead'' {\em or} ``nucleus excited; cat alive'', which prevents the cat to be dead {\em and} alive simultaneously, becomes possible only when environment-induced decoherence~\cite{Joos-03} destroys the quantum superpositions of such orthogonal realities.}

Finally, it is worth emphasizing that the possibility of accommodating the term IFM within the conceptual framework of some specific interpretation of the quantum theory is not discarded at all. Actually, this is explicitly observed by EV in their work~\cite{Elitzur-93}: ``The argument which claims that this is an interaction-free measurement sounds very persuasive but is, in fact, an artifact of a certain interpretation of quantum mechanics (the interpretation that is usually adopted in discussions of Wheeler's delayed-choice experiment). The paradox of obtaining information without interaction appears due to the assumption that only one ``branch'' of a quantum state exists.'' These words emphasize a point that seems to have recurrently been ignored so far: The term IFM is deeply interpretation dependent. 

\begin{acknowledgements}
The author thanks CNPq for financial support (process 471072/2007-9) and A.S. Sant'Anna for insightful discussions. Also, it is a pleasure to acknowledge one of the reviewers, whose suggestions helped to improve significantly the contents of this work. 
\end{acknowledgements}

 
\end{document}